\documentclass[5p]{elsarticle}

\usepackage[utf8]{inputenc}
\usepackage[T1]{fontenc}

\usepackage{latexsym}
\usepackage[dvipsnames]{xcolor}
\usepackage{braket}
\usepackage{amsmath}
\usepackage{amssymb}
\usepackage{csquotes}
\usepackage{bm}
\usepackage{graphicx}
\usepackage{float}
\usepackage{dcolumn}
\usepackage{mathtools}
\usepackage[version=3]{mhchem}
\usepackage{siunitx}

\usepackage{hyperref}
\usepackage{bookmark}

\def\Id{\ensuremath{\text{Id}}}
\def\dd{\ensuremath{\mathrm{d}}}
\def\ee{\ensuremath{\mathrm{e}}}
\def\ii{\ensuremath{\mathrm{i}}}

\def\gfo{\hat{\cal G}}
\def\gf{\mathcal{G}}

\def\energy{E}

\def\Tpasbar{T}

\usepackage{tikz}
\usetikzlibrary{calc}
\usetikzlibrary{decorations}

\hypersetup{
	colorlinks = true,
}
\AtBeginDocument{
\hypersetup{
	colorlinks = true,
	linkcolor  = Aquamarine!50!black,
	citecolor  = Aquamarine!50!black,
	urlcolor   = Aquamarine!50!black,
}
}

\newif\ifpdffigures
\pdffigurestrue

\ifpdffigures
	\relax
\else
	\usetikzlibrary{external}
\fi

\begin{document}

\title{Probing (topological) Floquet states through DC transport}

\author[ensl]{M. Fruchart}
    \ead{michel.fruchart@ens-lyon.fr}
\author[ensl]{P. Delplace}
    \ead{pierre.delplace@ens-lyon.fr}
\author[spsms]{J. Weston}
    \ead{joseph.weston@cea.fr}
\author[spsms]{X. Waintal}
    \ead{xavier.waintal@cea.fr}
\author[ensl]{D. Carpentier}
    \ead{david.carpentier@ens-lyon.fr}

\def\pgffig#1{
\ifpdffigures
	\includegraphics{figures-pdf/#1.pdf}
\else
	\bgroup
	\tikzsetnextfilename{#1}
	\input{figures/#1.pgf}
	\egroup
\fi
}
    
\address[ensl]{Laboratoire de Physique de l'École Normale Supérieure de Lyon, UMR CNRS 5672, 46 Allée d'Italie, 69007 Lyon, France}
\address[spsms]{CEA/Université Grenoble Alpes, INAC-SPSMS, F-38000 Grenoble, France}

\begin{abstract}
We consider the differential conductance of a periodically driven system connected to infinite electrodes.  We focus on the situation where the dissipation 
occurs predominantly in these electrodes. 
Using analytical arguments and a detailed numerical study we  relate the differential conductances of such a system in two and three terminal geometries 
to the spectrum of quasi-energies of the Floquet operator. 
 Moreover these differential conductances are found to provide an accurate probe of the existence of gaps in this quasi-energy spectrum, 
 being quantized when topological edge states occur within these gaps. Our analysis opens  the perspective to describe 
 the intermediate time dynamics of driven mesoscopic conductors as topological Floquet filters. 
\end{abstract}

\maketitle

\tableofcontents

\section{Introduction}

Recently, the possibility to  induce an out-of-equilibrium topological state of matter through irradiation or a periodic driving has stimulated numerous works. While initially 
 the external driving perturbation was used to trigger a  phase transition between states of conventional topological order \cite{Inoue:2010,Lindner:2011,Kitagawa:2011}, 
 fascinating topological properties specific to driven out-of-equilibrium states were soon identified \cite{Kitagawa:2010,Rudner:2013,Carpentier:2015}. 
While several proposals to realize and probe these topological states in various artificial systems have turned out to be successful 
\cite{Fang:2012,Kitagawa:2012,Hauke:2012,Rechtsman:2013,Hu:2015,Karzig:2014,Reichl:2014,Jotzu:2014}.
Their realization  in condensed matter have proved to be challenging \cite{Wang:2013,Onishi:2014}.

There is a strong analogy between equilibrium topological insulators and topological driven states. Both require the existence of a gap in 
the spectrum characterizing their single particle states: topological insulators are band insulators with a gap in the energy spectrum of the single particle 
Hamiltonian while topological driven states have a gap in the spectrum of the Floquet operator. 
In both cases,  a nontrivial topology manifests itself through the appearance within this gap of robust states located at the edge of the system. 
However, while in an insulator the gap separates empty states from occupied states, 
the thermodynamics of gapped periodically driven states is much less understood. Recent studies 
have stressed the differences between the nature of the states reached at long time in such periodically driven systems and the equilibrium 
ground states of insulators \cite{Lazarides:2014,Dehghani:2014,Lacedola:2015,Seetharam:2015}. 

Here we follow a different route: we focus on the relation between the DC transport of a periodically driven system and its quasi-energy Floquet 
spectrum in a regime where the times of flight of electrons through the system are shorter than the characteristic inelastic scattering times, which can be the case in mesoscopic systems. This provides a way to avoid the issue of long time dynamics of driven systems, which was raised in recent studies~\cite{Lacedola:2015,Seetharam:2015,Dehghani:2015}. 
Technically, this requires that the dominant perturbation of the unitary evolution of the driven system is the presence of the electrodes: \emph{dissipation 
should occur in the leads}. 
From this point of view, the driven system behaves as a topological Floquet filter instead of an out-of-equilibrium steady-state analogue of an equilibrium insulating state. 

DC transport is known to be an ideal  probe of the existence of edge states in topological equilibrium phases 
realized in condensed matter, particularly in two dimensions. 
In a seminal paper \cite{Buttiker:1988} Markus Büttiker demonstrated how the non-local conductances in a Hall bar fully characterize 
the nature of the quantum Hall effect and the associated chiral edge states. 
This approach was recently extended to study the quantum spin Hall effect occuring in HgTe/CdTe quantum wells \cite{Roth:2009,buttiker2009}. In this 
time-reversal invariant topological phase, the existence of a Kramers pair of counter-propagating edge states leads to a series of non-local conductances 
whose experimental observation clearly identified this new phase. 
For topological driven systems, the situation is more confusing: 
building on earlier works on the transport through a topological periodically driven state  \cite{Gu:2011,Kitagawa:2011}, recent studies have focused on 
the transport through a one-dimensional topological superconducting state \cite{Kundu:2013}, 
the effect on transport of the competition between heating by the drive and the coupling to the leads  \cite{Dehghani:2015} or 
the quantization of conductances of a topological phase in multi-terminal geometry~\cite{FoaTorres:2014}.
It was also proposed to probe quasienergy spectra (and topological edge states) through magnetization measurements \cite{dahlhaus2015} and tunneling spectroscopy \cite{fregoso2014}.
However, the relation between 
transport and the existence of topological edge states in periodically driven states remains unclear, and a summation procedure over different energies in 
the lead was proposed to recover a quantized conductance  \cite{Kundu:2013,Farrell:2015}.   
The purpose of our paper is to reconsider the relation between the (non-local) differential conductances of periodically driven systems and 
their Floquet quasi-energy spectrum, allowing for a direct relation between these differential conductances and the topological indices associated with the 
spectral gaps. In particular we will establish a protocol in a multi-terminal geometry allowing for this identification. 
In this point of view, a topological periodically driven system is viewed as a 
\emph{topological Floquet filter} with selective edge transport occurring for specific voltage biases between a lead and the system. These voltage biases lead to a stationnary DC current by counterbalancing the time dependence of Floquet states.

\section{From Floquet theory to scattering theory}

\subsection{Floquet theory for open systems}

We consider a periodically driven quantum system connected to $N_{\text{leads}}$ equilibrium electrodes through good contacts with large transmissions. 
The system is described by a Hamiltonian $\hat{H}^{\text{sys}}(t) - \hat{\Sigma} $ where 
$\hat{H}^{\text{sys}}(t+T)=\hat{H}^{\text{sys}}(t)$ with $T$  the period of the drive, and $ \hat{\Sigma}$ is a self-energy accounting for the 
coupling between the system and its environment (e.g. the leads). We assume in the following that this self-energy is dominated by the exchange with the 
electrons in the leads. When all characteristic times of the leads are small with respect to the characteristic times of the system, we can use the so-called wide band approximation \cite{Kohler:2005} where the self-energy is assumed to be constant in energy: $\hat{\Sigma}(E) \simeq \hat{\Sigma}$.
The dynamics of the system is described by the evolution operator $\hat{U}(t,t')$ which obeys the equation 
\begin{equation}
	\ii \hbar \frac{\dd}{\dd t}\hat{U}(t,t') = \left( \hat{H}^{\text{sys}}(t) -  \hat{\Sigma} \right) \hat{U}(t,t') .
\label{eq:schrodinger_U}
\end{equation}
Of great importance is the \textit{Floquet operator} which is the evolution operator after one period $\hat{U}(T,0)$. 
When diagonalizable, it can be decomposed on the left eigenstates  $\bra{\tilde{\phi}_\alpha}$  and the right eigenstates  $\ket{\tilde{\phi}_\alpha}$ of $\hat{U}(T,0)$
\begin{align}
 \begin{split}
	\hat{U}(T,0) \ket{\phi_\alpha} &= \lambda_\alpha \ket{\phi_\alpha} , \\
	\bra{\tilde{\phi}_\alpha}\hat{U}(T,0) &= \lambda_\alpha \bra{\tilde{\phi}_\alpha} , 
\end{split}
\label{eq:lambda}
\end{align}
that constitute a bi-orthonormal basis of the Hilbert space
\begin{equation}
	\braket{\tilde{\phi}_\alpha \mid \phi_\beta} = \delta_{\alpha \beta} \ ; \ 
	\sum_{\alpha} \ket{\phi_\alpha}\bra{\tilde{\phi}_\alpha} = \Id .
\label{eq:basis}
\end{equation}
The eigenvalues $\lambda_\alpha$ in Eq.~\eqref{eq:lambda} are called the Floquet multiplicators and read
\begin{equation}
	\lambda_\alpha = \exp\left[ - \ii \left( \frac{\varepsilon_\alpha}{\hbar}  - \ii \gamma_\alpha \right) T \right].
\end{equation}
The coefficient $ \varepsilon_{\alpha}$ is called the quasienergy and $\gamma_{\alpha}$ is its damping rate whose inverse gives the life-time of the eigenstate.
Note that the quasienergy being a phase, it is defined modulo the driving frequency $\omega=2\pi/T$.
Any state at arbitrary time $t$ can then be constructed from the eigenstates of the Floquet operator. 
It is particularly useful to define the left and right Floquet states
\begin{align}
 \begin{split}
	\ket{u_\alpha(t)} &= \ee^{ \ii \left( \varepsilon_\alpha/\hbar  - \ii \gamma_\alpha \right) t} \   \hat{U}(t,0)\ket{\phi_\alpha}  , \\
	\bra{\tilde{u}_\alpha(t)} &= \ee^{- \ii \left( \varepsilon_\alpha/\hbar  - \ii \gamma_\alpha \right) t}\ \bra{\tilde{\phi}_\alpha}  \hat{U}(0,t) , 
\end{split}
\label{eq:u_def}
\end{align}
which are periodic in time, $\ket{u_\alpha(t)}=\ket{u_\alpha(t+T)}$ (same for $\bra{\tilde{u}_\alpha(t)}$), so that they can be expanded in Fourier series
\begin{align}
 \begin{split}
	\ket{u_{\alpha}(t)} &= \sum_{p \in \mathbb{Z}} \ee^{-\ii p \omega t} \ket{u_{\alpha}^{(p)}} , \\
	\bra{\tilde{u}_{\alpha}(t)} &= \sum_{p \in \mathbb{Z}} \ee^{\ii p \omega t} \bra{\tilde{u}_{\alpha}^{(p)}} , 
\label{eq:u_fourier}
\end{split}
\end{align}
where the harmonics read
\begin{align}
 \begin{split}
 \ket{u_{\alpha}^{(p)}} = \frac{1}{T} \int_{0}^{T} \dd t \; \ee^{\ii p \omega t} \ket{u_{\alpha}(t)} \\
 \bra{\tilde{u}_{\alpha}^{(p)}} = \frac{1}{T} \int_{0}^{T} \dd t \; \ee^{-\ii p \omega t} \bra{\tilde{u}_{\alpha}(t)} .
\end{split}
\label{eq:harmonics}
\end{align}
From  Eqs. (\ref{eq:basis}) and (\ref{eq:u_def})  the evolution operator can be expanded on the Floquet states as
\begin{equation}
\hat{U}(t, t') = \sum_{\alpha} \ee^{-\ii \left( \varepsilon_\alpha/\hbar  - \ii \gamma_\alpha  \right)\left(t-t' \right)} \ket{u_{\alpha}(t)}\bra{\tilde{u}_{\alpha}(t')} .
\end{equation}
This expression can finally be decomposed on the harmonics of the Floquet states by using Eq.(\ref{eq:u_fourier})
\begin{multline}
	\label{eq:U_decomposition}
	\!\!\!\!\hat{U}(t, t') = \!\!\!
	\sum_{\substack{\alpha \\ p,p' \in \mathbb{Z}}} \!\!\! \ee^{-\ii \left[ \left( \frac{\varepsilon_\alpha}{\hbar}  - \ii \gamma_\alpha  \right)(t-t') + \omega (p t - p' t')\right]}
	\ket{u_{\alpha}^{(p)}}\!\bra{\tilde{u}_{\alpha}^{(p')}}.
\end{multline}

In practice, the spectrum of Floquet operator of the semi-infinite system can be obtained numerically either by direct a computation of $U(T,0)$ (e.g. as a discretized in time version of the infinite product) or through its representation in Sambe space \cite{sambe73}.

\subsection{Differential conductance}

Based on a standard formalism, we can calculate analytically the differential conductance of the periodically driven system in a multi-terminal geometry
and relate it to the quasienergy spectrum of the system.
We follow the standard Floquet scattering formalism  \cite{Moskalets:2002,Kohler:2005,Stefanucci:2008,Gaury:2013,Kundu:2013,FoaTorres:2014} to describe the 
transport properties of this multiterminal setup in a phase coherent regime. 
We consider the rolling average over a period~$T$ (all time-averages in the following are also rolling averages over one driving perdiod) of the current entering each lead labelled by the index~$\ell$:
\begin{equation}
	\mathcal{I}_{\ell}(t) =  \frac{1}{T} \int_{t}^{t+T} \dd t' \; \braket{\hat{J}_{\ell}(t')}.
	\label{eq:averagedCurrent}
\end{equation}
where $\braket{\hat{J}_{\ell}(t')}$ is the expectation value of the current entering lead~$\ell$ at time~$t'$.
This average current satisfies a relation~\cite{Moskalets:2002,Kohler:2005,Stefanucci:2008}:
\begin{equation}
	\mathcal{I}_{\ell}(t) = \frac{e}{h}  \int \dd\energy\; 	
 	\sum_{\ell' \neq \ell} \left[
	 \Tpasbar_{\ell \ell'}(t, \energy) f_{\ell'}(\energy)
	  - \Tpasbar_{\ell' \ell}(t, \energy) f_{\ell}(\energy)
	\right], 
	\label{eq:Il_stat}
\end{equation}
where $f_{\ell}(\energy)$ is the Fermi-Dirac distribution of the lead $\ell$ assumed to be at equilibrium at the chemical potential $\mu_{\ell}$. 
The $\Tpasbar_{\ell \ell'}(\energy) $ are the time-averaged transmission coefficients between lead $\ell$ and $\ell'$ which will be discussed below.  
We define the differential conductance $G_{\ell \ell'}(t, E)$ as the sensitivity of the current entering the lead $\ell$ to variations of the chemical potential $\mu_{\ell'}$ 
of the lead\footnote{See \cite{Gaury:2013} for a recent discussion  of chemical versus electrical potential drops at the interface between the system and an electrode. Such 
subtleties are quietly neglected in the following.} $\ell'$ : 
\begin{equation}
        G_{\ell \ell'} (t, E)\equiv - e \left. \frac{\dd \mathcal{I}_{\ell}}{\dd \mu_{\ell'}}  \right|_{\mu_{\ell'}=\mu_{\text{sys}}+E}.
	\label{eq:diff_cond}
\end{equation}
Note that this definition is not symmetric in the various chemical potentials $\mu_{\ell}$, as opposed to other definitions used in the literature. 
In the long time stationary regime on which we focus, $t \to \infty$, the average conductances $G_{\ell \ell'}(t,E)$ are expected to reach a value independent on the choice of origin of time~$t$ and associated initial conditions \cite{Kohler:2005}, which we denote by $G_{\ell \ell'}^{\infty}(E)$.

We obtain from Eq.~\eqref{eq:Il_stat} the zero temperature time-averaged differential conductances 
\begin{align}
        G_{\ell \ell'}^{\infty}(E) &= \frac{e^2}{h}  \Tpasbar_{\ell \ell' }(E) \qquad \text{for $\ell \neq \ell'$}
	\label{eq:diff_condllp}
\\
        G_{\ell \ell}^{\infty}(E) &= - \frac{e^2}{h} \sum_{\ell' \neq \ell}  \Tpasbar_{\ell' \ell }(E), 
	\label{eq:diff_condll}
\end{align}
which satisfy the rule $\sum_{\ell} G_{\ell ,\ell'}(E) = 0$  for any $ \ell'$ (the average current leaving the system is independent on the chemical potentials in the leads).   
This formula is analogous to the Landauer-B\"uttiker formula for 
the differential conductance of multiterminal equilibrium systems. 

\subsection{Generalized Fisher-Lee relations}

The average transmission coefficients $\Tpasbar_{\ell \ell'}(\energy)$ can be related to the Floquet-Green functions of the system, in a way analogous to the 
case of undriven conductors~\cite{Fisher:1981}. 
The retarded Green functions in mixed energy-time representations are defined by 
\begin{equation}
{\hat{\gf}}(t,E) = \frac{1}{\ii \hbar}\int d\tau \ e^{\ii E\tau/\hbar} \,  \gfo(t,t-\tau) 
\end{equation}
where $\gfo(t,t')$ satisfies
\begin{align}
 \begin{split}
\label{eq:def_gf}
	\left( \ii \hbar \frac{\dd}{\dd t} - \hat{H}^{\text{sys}}(t)  + \hat{\Sigma} \right) \gfo(t,t') = \delta(t-t').
 \end{split}
\end{align}
 It is  related to the evolution operator defined in Eq.(\ref{eq:schrodinger_U}) by
\begin{equation}
 {\gfo}(t,t')=\frac{1}{\ii \hbar}\Theta(t-t')\hat{U}(t,t') .
\end{equation}
From the decomposition Eq.(\ref{eq:U_decomposition}) of $\hat{U}(t,t')$ over the harmonics of the Floquet states, 
we  express the \textit{Floquet-Green functions}
\begin{equation}
	\label{eq_green_floquet}
	\hat{\gf}^{(p)}(\energy) = \frac{1}{T} \int_{0}^{T} \dd t\; \ee^{\ii p \omega t} \, {\gf}(t, \energy) .
\end{equation}
as 
\begin{equation}
	\label{eq_green_floquet_evolution}
	\hat{\gf}^{(p)}(\energy) =
	 \sum_{r, \alpha} \frac{\ket{u_{\alpha}^{(p+r)}}\bra{\tilde{u}_{\alpha}^{(r)}} }{\energy - \left[ \varepsilon_\alpha + r \hbar \omega - \ii \hbar \gamma_\alpha \right]}.
\end{equation}

The transmission coefficients are expressed as 
\begin{equation}
	\Tpasbar_{\ell' \ell}(\energy)	= \sum_{p \in \mathbb{Z}} T_{\ell' \ell}^{(p)}(\energy) , 
	\label{eq:Tbar}
\end{equation}
where each  $T^{(p)}_{\ell' \ell }(E)$ is the transmission coefficient for an electron injected in lead $\ell$ at the energy $E=\mu_{\ell}$ and 
leaving the system in lead $\ell'$ at the energy $E+p \hbar \omega$, {\it i.e.}  after having  exchanged $p$ quanta with the driving perturbation. 
They read  (see \ref{sec:appendix} and \cite{Kohler:2005,Stefanucci:2008,Gaury:2013,arrachea2006}): 
\begin{equation}
	\label{eq:coefficient_transmission_general}
	T_{\ell' \ell}^{(p)}(\energy) = \text{Tr} 
	\left[ \hat{\gf}^{{\dagger}(p)}(\energy) \hat{\Gamma}_{\ell'}(\energy + p \hbar \omega) \hat{\gf}^{(p)}(\energy) \hat{\Gamma}_{\ell}(\energy) \right]
\end{equation}
where $\hat{\Gamma}_{\ell}(\energy)$ is the coupling operator at energy $E$ between the system and the electrode $\ell$. 

Note that when the rates $\gamma_\alpha$ are sufficiently small for the quasi-energies $\epsilon_\alpha$ occurring in Eq.~\eqref{eq_green_floquet_evolution} 
to be \enquote{well-defined}, the equations (\ref{eq_green_floquet_evolution},\ref{eq:Tbar},\ref{eq:coefficient_transmission_general}) imply that the 
transmission coefficient $\Tpasbar_{\ell' \ell}(\energy))$ and thus the differential conductance $G_{\ell' \ell}(E)$ vanishes whenever the 
energy $\mu_{\ell}=E$ does not correspond to a quasi-energy $\epsilon_\alpha$ up to a multiple of~$\hbar \omega$.  This is nothing but the conservation of energy 
of incident state which holds modulo $\hbar \omega$ in a periodically driven system. 
 This property allows one to probe the existence of gaps in the quasi-energy spectrum of the Floquet operator. 
  Note however that to access the whole quasi-energy spectrum, the lead $\mu_{\ell}$ has to be strongly biased (with a bias of order~$\hbar \omega$) with respect to the 
  scattering region at $\mu_{\text{sys}}$, a situation far from the standard procedure to probe equilibrium phases, but required to probe an inherently out-of-equilibrium 
  Floquet state.  
Moreover, in the presence of a ballistic mode connecting two leads $\ell$ and $\ell'$ such as the edge mode of a topological Floquet state, 
 we expect the corresponding $\Tpasbar_{\ell' \ell}(\energy)$ to be quantized \emph{provided both interfaces are sufficiently transparent}. Note however 
 that as the states leaving the systems have component of energies $E+p\hbar \omega$, this implies a perfect transparency over a broad range of energies. 
 These two properties allow for a potential probe of topological Floquet states through non-local transport. 
 In the next section, we study this application by a numerical implementation of transport through an AC driven system.

\section{Numerical study}

In the following, we provide numerical evidence of the correspondence between the differential conductance and the quasi-energy spectrum in a periodically driven system. In particular, we show that a single out-of-equilibrium topological edge state corresponds to a quantized differential conductance. 
The chirality of these topological edge states is probed in a multi-terminal setup. 

\subsection{Model and method}
\begin{figure}
	\centering
	\input{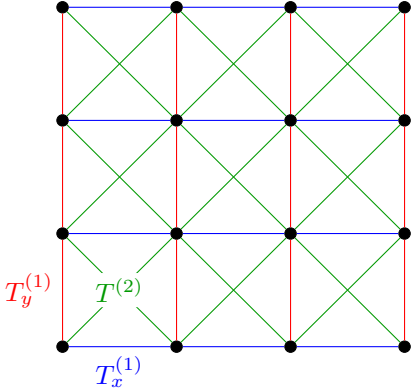}
	\vspace*{-0.25cm}
	\caption{\label{fig:reseau} Lattice of couplings for the half-BHZ model with two orbitals per site of the square lattice. 
	Blue and red links correspond to the $ T^{(1)}_x$ and $ T^{(1)}_y$ nearest neighbor 
	coupling matrices between $s$ and $p$ orbitals while the green links represent  $T^{(2)}$ the second-nearest neighbor coupling matrix. }
\end{figure}
\begin{figure*}[t]
\centerline{
	\input{figures/fig_qe_spectrum_conductance_trivial.pgf}
	\input{figures/fig_qe_spectrum_conductance_topological.pgf}
	}
	\vspace*{-0.25cm}
	\caption{\label{fig_qe_spectrum}
	Quasi-energy spectrum of the driven  half-BHZ model in an infinitely long ribbon and corresponding asymptotic two terminal differential conductance
	$G_{\text{RL}}^{\infty}(E)$ for  finite system connected to infinite leads. The results are shown for parameters with only trivial gaps with driving amplitude $F=2$ 
	(Left) and a topological gap at $\epsilon=0$ or $F=8$  (Right). 
	The quasi-energy spectra are obtained for strip of width $W=\num{60}$ sites by diagonalization in Sambe space with~\num{5} sidebands. 
	The quasi-energies $\varepsilon$  are shown in units of the driving frequency $\hbar \omega$ as a function of the quasi-momentum in the direction 
	of the ribbon $k_x$. 
	For the topological case ($F = \num{8.}$, Right figure), states appear inside the gap $\epsilon=0$, located on either side of the ribbon. 
	The location of each Floquet mode was numerically computed and converted to a color using the color scale of figure \ref{fig:figure_geometry}. Delocalized states (and states located in the bulk) are in green, whereas states located on an edge are in red or blue, depending on which edge. This signals the appearance of a topologically non-trivial gap around $\epsilon=0$. On the other hand, the gaps in 
	$\epsilon= \pm \hbar \omega /2$ remains topologically trivial in both cases. 
	The asymptotic long time limit of the corresponding two-terminal differential conductance is displayed as a function of  the chemical potential of the left lead 
	(black dots and lines) for a finite sample of size $W\times L=\num{60}\times\num{30}$ sites. 
	This differential conductance $G_{\text{RL}}^{\infty}$ is quantized, with $G = 1 e^2/h$ in the topological quasi-energy gap (Right) and vanishes in the trivial gap (Left). 
	Hence it is an accurate probe of the existence of topological gap in the quasi-energy spectrum of open conductors connected to electrodes. 
	The colored arrows indicate the chemical potential of the incoming lead used in the computations of Fig.~\ref{fig_conductance_time_short}.
	}	
\end{figure*}
\begin{figure}[hbt]
	\centering
	\input{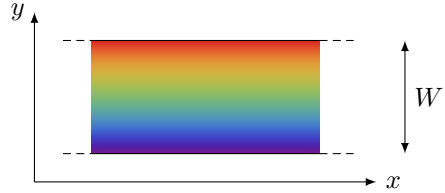}
	\caption{\label{fig:figure_geometry} Geometry of the semi-infinite system used to compute quasi-energy spectra of figure \ref{fig_qe_spectrum}. In those spectra, the color of each point represents the average $y$ position of the corresponding Floquet mode, according to the color scale shown on this figure.  }
\end{figure}
Following \cite{Rudner:2013}, we use the restriction to spins up of the Bernevig-Hugues-Zhang model of 
quantum spin Hall effect in \ce{HgTe}/\ce{CdTe} quantum wells~\cite{Bernevig:2006} (referred to as half-BHZ model). As the Haldane model~\cite{Haldane:1988} it realizes an anomalous quantum Hall equilibrium phase, but on a square lattice 
with  two orbitals per site, denoted $s$ and $p$. 
The tight-binding Hamiltonian with nearest and next-to-nearest neighbors hoppings (see Fig.~\ref{fig:reseau}) 
 can therefore be written as a two-by-two matrix on the $(s,p)$ basis as 
\begin{equation}
\begin{split}
	&H_{\text{BHZ}} = \sum_{x,y} \biggl[  \left[\left(M - J - 4 B \right) \sigma_z - \mu \sigma_0 \right] \ket{x,y} \bra{x,y} \\
	&+T^{(1)}_x  \ket{x,y} \bra{x+1,y} +T^{(1)}_y \ket{x,y} \bra{x,y+1} \\
	&+T^{(2)} \left( \ket{x,y}  \bra{x+1,y+1} + \ket{x,y}  \bra{x+1,y-1} \right)	\biggr] \\
	&+ \text{h.c.}
\end{split}
\end{equation}
with hopping matrices 
\begin{equation}
\begin{split}
	T^{(1)}_{x/y} = \ \frac{A}{2 \ii} \sigma_{x/y} + B \sigma_z
	\text{ and }
	T^{(2)} =  \frac{J}{4} \sigma_z
\end{split}
\end{equation}
where $\sigma_{x,y,z}$ are the Pauli matrices and $\sigma_0$ the identity matrix. The parameters of this Hamiltonian are chosen as~$M=\num{-1.}$, $J = \num{1.5}$, $A = \num{4.}$, $B = \num{1.5}$ : they correspond to an equilibrium phase which is a trivial insulator.  A topological insulating phase can be reached by 
varying {\it e.g.} $M$ such that $0<M/B<4$ or $4<M/B<8$. 
This trivial equilibrium phases is submitted to a periodic on-site perturbation \cite{Rudner:2013} 
\begin{equation}
	\Delta H(t) = \sum_{x,y} F \left[  \sin(\omega t) \sigma_x +  \cos(\omega t) \sigma_y \right] \ket{x,y} \bra{x,y}. 
\end{equation}
Note that this perturbation does not correspond to the variation of a parameter of the initial Hamiltonian.
Throughout our study, we have used a driving frequency $\omega = \num{20}$. 
Depending on the strength $F$, 
this perturbation can drive the system either towards a topologically nontrivial out-of-equilibrium state or in a topologically trivial out-of-equilibrium state. 
 We have chosen two values $F=2$ and $F=8$ of the driving amplitude so that  gaps open at $\epsilon= 0, \pm \hbar \omega/2$ of the
  quasi-energy spectrum.  For each value of $F$ and each gap, 
 we have computed numerically the bulk topological invariant~$W_{\varepsilon}[U]$ associated to the quasi-energy gap around 
 $\epsilon$~\cite{Rudner:2013} to ensure that $F=2$ correspond to a trivial gapped Floquet state, while $F=8$ corresponds to a topological  gapped Floquet state
  with a non-trivial gap at $\epsilon= 0$.  Alternatively,  in the two cases 
we have computed the quasi-energy spectrum for the driven model in an infinitely long ribbon of width $W=\num{60}$ sites (see Fig.~\ref{fig:figure_geometry}) 
by diagonalization in Sambe space with~\num{5} sidebands. The resulting spectra are shown in Fig.~\ref{fig_qe_spectrum}: they show as expected that states located at each edge on the ribbon appear 
 inside the $\epsilon=0$ gap for the topological case ($F=8$) as opposed to the trivial case~($F=2$).

Finally, to study transport through the system, leads are attached to a finite size system. 
These leads are modeled by a simple tight-binding Hamiltonian on a square lattice with nearest neignbors hoppings with amplitude~$J_0=\num{8}$. An onsite potential is added to the lead Hamiltonian to reduce mismatches between the incoming and outgoing states of the leads and the scattering states of the central region.

The numerical calculations are performed using the numerical method described in \cite{Gaury:2013}. 
 Although the technique is based on wavefunctions, it is mathematically equivalent to the Green function approach used in this article
(see Eq.~(49) of \cite{Gaury:2013} for the connection with the transmission coefficient as well as section 5.4 for the link with Floquet theory). Our implementation is based on the Kwant package \cite{Groth:2014}.

\begin{figure*}[!ht]
\centerline{
	\input{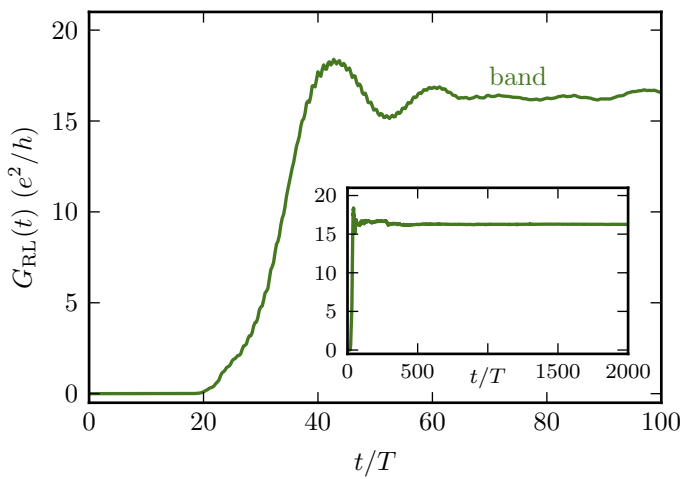}\hspace*{-0.75cm}
	\input{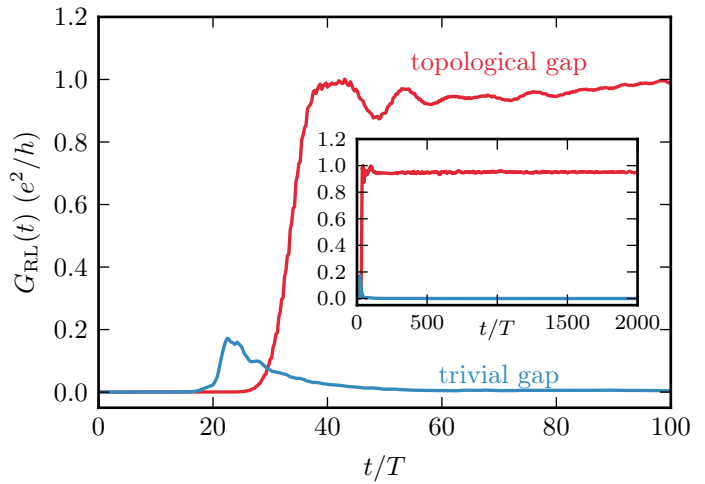}
	}
	\vspace*{-0.25cm}
	\caption{\label{fig_conductance_time_short} Two-terminal differential conductance as a function of time, in the transient regime, for two chemical potentials, in a topologically trivial case. Green curve: incoming chemical potential in a quasi-energy band ; blue and red curves: incoming chemical potential in the bulk quasi-energy gap, in a trivial (blue) and a topological (red) case. Inset: longer simulations were carried out to ensure convergence. The asymptotic value is denoted by~$G_{\text{RL}}^{\infty}$.}
\end{figure*}
\subsection{Probing the quasi-energy bands and gaps through two-terminal differential conductance}

The differential conductance $G_{\text{RL}}(t)$ is computed in a two-terminal setup through a sample of width $W=60$ and length $L=\num{30}$ sites. 
Throughout this study, the chemical potential of the undriven system was chosen as~$\mu_{\text{sys}}=0$. 
To get rid of oscillations faster than the driving frequency, which are irrelevant for our purpose, we perform a sliding average over one driving period~$T$. 
After a transient regime, the (averaged) differential conductance~$\bar{G}(t)$ converges to a finite value (see Fig.~\ref{fig_conductance_time_short}). 
This transient regime can be understood as the time of flight of the state injected at the left lead to the right lead after the driving perturbation has been turned 
on and the Floquet states developed inside the system. When the chemical potential of the incoming lead lies in a topological quasi-energy gap of the scatterng region, transport occurs through a chiral state localized near the edge of the sample. 
 We can easily evaluate the travel length $L$. The expected group velocity is 
 extracted from the slope of the quasi-energy dispersion relation from Fig.~\ref{fig_qe_spectrum} through 
\begin{equation}
	v_{\text{g}} = \frac{1}{\hbar} \frac{\dd \varepsilon}{\dd k}, 
\end{equation}
and we obtain $v_{\text{g}} = \num{0.15(1)} \, a \omega$ where $a$ is the lattice spacing and $\omega$ the driving  frequency. This correlates 
perfectly with the time~$\Delta t$ of the first increase from zero of the conductance from the switching on of the driving field. From 
 the curve on Fig.~\ref{fig_conductance_time_short} we find  find~$L/\Delta t= \num{0.14(1)} \, a \omega$.

\begin{figure}[!t]
	\centering
	\input{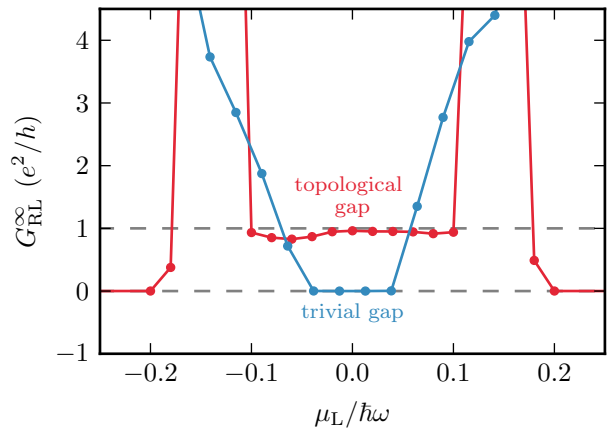}
	\vspace*{-0.5cm}
	\caption{\label{fig_conductance_bias_zoom} Zoom of Fig.~\ref{fig_qe_spectrum} : asymptotic two-terminal differential conductance $G_{\text{RL}}^{\infty}$ in units of $e^{2}/h$ 
	with respect to the chemical potential of the left lead (black dots and lines) 
	around the quasi-energy gap $\epsilon=0$ for a trivial (red) and topological gap (red). 
	The reference energy $\mu_{\text{L}}=0$ corresponds to an unbiased lead whose chemical potential identifies with the one of the driven system 
	$\mu_{\text{sys}}=0$. 	
	The differential conductance vanishes in the quasi-energy gaps for a trivial spectral gap, but reaches an approximately constant and quantized value $G =  e^2/h$  
	within the topological gap with one chiral edge mode. 
	Deviations from $G_{\text{RL}}^{\infty} = 1 \, e^2/h$ are attributed to the finite dispersion relation of the leads, which does not completely satisfy the wide band approximation and leads to imperfect transparencies of the contacts.
	}
\end{figure}

After this transient regime, the differential conductance $G_{\text{RL}}(t)$ reaches a long time stationary limit~$G_{\text{RL}}^{\infty}$, as shown in the inset of 
 Fig.~\ref{fig_conductance_time_short}. As expected, this asymptotic differential conductance is 
   sensitive to the quasi-energy spectrum of the driven system: when in a spectral band, it reaches high values but 
vanishes when the chemical potential $\mu_{\text{L}}- \mu_{\text{sys}}$ lies in a trivial spectral gap of the Floquet operator, as shown on the left plot of 
Fig.~\ref{fig_conductance_time_short}. Moreover, when this spectral gap is topological, the associated presence of chiral states at the edge of the system 
shown in Fig.~\ref{fig_qe_spectrum} (Right) leads to a quantized two terminal conductance as shown in the right plot of Fig.~\ref{fig_conductance_time_short}. 
 To further study the correlation between the differential conductance as a function of the chemical potential $\mu_{\text{L}}- \mu_{\text{sys}}=E$ and 
 the spectral gap of the system we have studied the behavior of the long time limit of this conductance as a function of~$E$.  The corresponding results as well the 
  spectra of the isolated driven models are plotted in Fig.~\ref{fig_qe_spectrum} for both the trivial and a topological gapped Floquet states. 
 We find a strong correlation between the vanishing of both quantities for the topologically trivial case: the differential conductance vanishes only 
 inside a trivial gap,  except at the edge of the gap where finite size effects occurs due to imperfect transparencies of the contact with the leads, 
 as shown on a magnified view around the gap $\epsilon=0$ in Fig.~\ref{fig_conductance_bias_zoom}. This demonstrates both that the quasi-energy spectrum 
 of the finite system connected to infinite leads is sufficiently close to the spectrum of the isolated infinite strip, and that the differential conductance is an accurate 
 probe of this spectrum for the open system. 
Moreover, in the topological case the asymptotic differential conductance remains constant equal to the number $n=1$ of edge states (in units of~$e^2/h$) inside the topological gap 
as shown in Fig.~\ref{fig_conductance_bias_zoom}. The small deviations from $G_{\text{RL}}^{\infty} = 1 \, e^2/h$ visible in figure \ref{fig_conductance_bias_zoom} are attributed to the finite dispersion relation of the leads, which does not completely satisfy the wide band approximation and leads to imperfect transparencies of the contacts.

\subsection{Multiterminal geometry}

\begin{figure}[!ht]
	\input{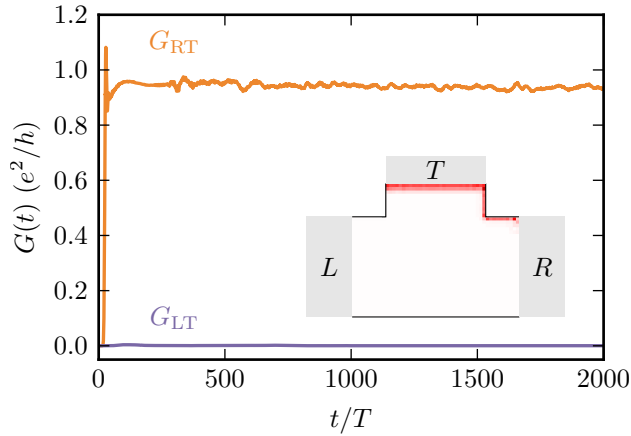}
	\vspace*{-0.5cm}
	\caption{\label{fig_conductance_time_three_terminal_bands} Three-terminal differential conductance $G_{\text{LT}}$ (violet) and 
	$G_{\text{RT}}$ (orange) as a function of time for a chemical potential of the top lead in a topological bulk gap. 
	After an initial transient regime,  asymptotic values for the nonlocal conductances are reached with 
	$G_{\text{LT}}^{\infty} = \num{0.0002(1)} \, e^2/h$ and~$G_{\text{RT}}^{\infty} = \num{0.94(1)} \, e^2/h$. 
	The simultaneous vanishing of $G_{\text{LT}}^{\infty}$ and quantized $G_{\text{RT}}^{\infty} $ to the two-terminal value  
	demonstrates the chiral nature of topological edge states. 
	This is further illustrated in the inset by the corresponding probability density $|\psi(t, x)|^2$ in the stationnary regime (here at $t =1000\ T$) which 
	is entirely localized around the $T\to R$ edge, demonstrating furthermore the high transparency of the $R$ contact for this edge mode. 	
	Simulations are done with a width of the contacts $W=\num{30}$ sites and a total length of \num{50} sites.
	}
\end{figure}

 A crucial characteristics of the edge states occurring in the topological gap of the spectrum shown in Fig.~\ref{fig_qe_spectrum} (Right) 
 is their chiral nature. 
 In the equilibrium case this property together with their ballistic propagation leads to quantized conductance in a Hall bar geometry  \cite{Haldane:1988}. 
 To test the chirality of the topological edge states, we have computed differential conductances 
  in a three-terminal geometry shown in Fig. ~\ref{fig_conductance_time_three_terminal_bands}, where the width of the contact with the electrodes 
  is $W=\num{30}$ sites, the total length (between $L$ and $R$ contacts) is $\num{50}$ sites, corresponds to all three arms having a length of $\num{10}$ sites. 
 In this geometry, we monitor the two differential conductances  
 \begin{equation}
 \begin{split}
     G_{R,T} (E) =  - e \left. \frac{\dd \mathcal{I}_{R}}{\dd \mu_{T}}  \right|_{\mu_{T}=\mu_{\text{sys}}+E} , \\
     G_{L,T} (E) =  - e \left. \frac{\dd \mathcal{I}_{L}}{\dd \mu_{T}}  \right|_{\mu_{T}=\mu_{\text{sys}}+E} . 
\end{split}
 \end{equation} 
 where $\mathcal{I}_{R,L}$ are the current in the $R,L$ contacts averaged over one period $T$ of drive (see eq.~\ref{eq:averagedCurrent}). 
 The chemical potential of the system is still set to~$\mu_{\text{sys}}=0$ (as if imposed e.g. by a backgate). 

\begin{figure}[!ht]
	\input{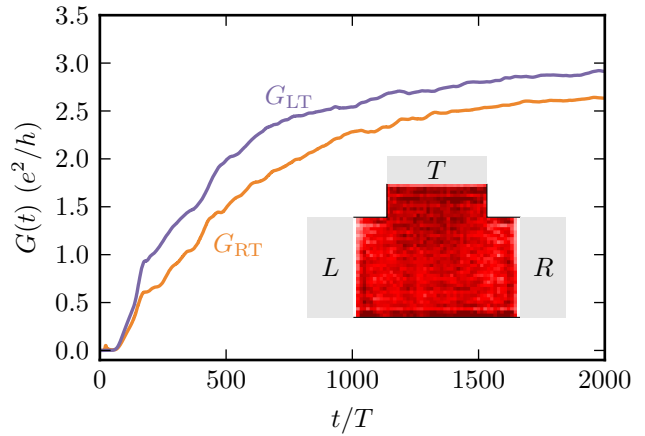}
	\vspace*{-0.5cm}
	\caption{\label{fig_conductance_time_three_terminal_bands_bulk} 
	Three-terminal differential conductance $G_{\text{LT}}$ (violet) and 
	$G_{\text{RT}}$ (orange) as a function of time for a chemical potential of the top lead in a quasi-energy band. A longer transient regime is observed, in agreement with 
	smaller group velocities than for the edge states, and large asymptotic values for both $G_{\text{LT}}$ and $G_{\text{RT}}$ are reached, confirming the 
	non-chiral nature of the corresponding Floquet states. 
	The inset shows color map representing the probability density $|\psi(t, x)|^2$ for at $t/T=1800$, demonstrating the delocalized nature of the Floquet states.
	}
\end{figure}

We consider the case of a topological gap round~$\epsilon=0$ (see Fig.~\ref{fig_qe_spectrum} (Right)).  
First we set the chemical potential of the top lead inside this gap ($\mu_{T}=0$).  
The time evolution of the differential conductances are shown in Fig.~\ref{fig_conductance_time_three_terminal_bands}.  After a transient regime, the differential conductances 
converge to asymptotic values $G_{\text{LT}}^{\infty} = \num{0.0002(1)} \, e^2/h$ and~$G_{\text{RT}}^{\infty} = \num{0.94(1)} \, e^2/h$. The value of 
$G_{\text{LT}}^{\infty}$ is in agreement with the two terminal results, while the vanishing of $G_{\text{RT}}^{\infty}$ is in perfect 
agreement with the chiral nature of the edge state moving clockwise for the chosen parameters. This can be contrasted with the  
case of a chemical potential $\mu_{T}=0$ set inside a bulk Floquet band, displayed in Fig.~\ref{fig_conductance_time_three_terminal_bands_bulk}. 
In this case, after a longer transient regime due to slower group velocities, both conductances converge towards large finite values, confirming the absence of chirality for 
these bulk states. 

Finally, to illustrate the spatial structure of these Floquet states we show in insets of Fig.~\ref{fig_conductance_time_three_terminal_bands} and~\ref{fig_conductance_time_three_terminal_bands_bulk} a color map of the probability density 
of the states reached at large times for both the topological gap and inside the bulk bands. The chiral nature of the states injected in the top lead is clearly apparent 
and confirms the transport results.

\section{Discussion}

We have studied the differential conductances $G_{\ell' \ell} (E)$ defined in Eq.~\eqref{eq:diff_cond} for both a two-terminal and a three-terminal geometry. 
As for equilibrium phases, 
we have found that these conductances probe the topological nature of a gapped Floquet state: it 
vanishes whenever the chemical potential $\mu_{\ell}$ lies in a topologically trivial gap while we found it to be quantized  
for chemical potential  inside a topological gap of quasi-energies. Moreover, the chiral nature of the edge states reflected on the value of the multi-terminal 
differential conductances within this gap. 
 These results validate the relation between the DC transport through a periodically driven system and the Floquet description of the dynamics of the isolated 
 system. This requires that the dissipation occurs mostly inside the leads, {\it i.e.} that the system is small enough for the travel times through it to be small compared with 
 times scales of other source of dissipation (phonons, photons, etc). In that situation, 
  the differential conductance of the system depends on the intermediate time evolution of the driven system and can be accurately described starting from the unitary 
 dynamics of the driven system and its associated topological properties \cite{Rudner:2013,Carpentier:2015}. 
 This should be the case as long as phase coherence is preserved on the length scale of the sample, even in the presence of dissipation or weak interactions.
  This result paves the way towards a direct engineering 
 of a topological filter through the implementation of Floquet states. 
 
 Our results on the quantization of differential conductances inside a topological gap contrast with some of the recent results on similar systems \cite{Kundu:2013,FoaTorres:2014,Farrell:2015}, where a summation procedure over 
 multiple chemical potentials was required to recover a quantized conductance. Note that these previous results used a different measuring scheme of differential 
 conductance, symmetric in chemical potentials of electrodes, which may filter out some of the energies of the outgoing electronic states, and thus require 
 a summation protocol. In our case, a strong and asymmetric voltage bias is used to compensate the time dependence (through quasi-energies) of the Floquet states.
 Moreover, we have used in our numerical study a model with a single edge mode located around $k\simeq 0$ as opposed to double-mode topological gaps
  used previously.  More complex edge structures may well lead to lower transparency of the interfaces for energies in the gap, and thus non-quantized conductances. 
  The study of the transparency of such interface between a strongly driven conductor and a DC electrode is a deep and topical subject whose 
  discussion goes beyond the scope of the present study.

{\bf Acknowledgments:} This paper is dedicated to the memory of Markus Büttiker whose pioneering works on mesoscopic systems and in particular on scattering theory of chiral edge channels shaped our understanding of electronic transport. P.D. will always be grateful to him for his trust, enthusiasm and advice he received during his stay as a post-doc assistant in the Büttiker's group. 
This work was supported by the French Agence Nationale de la Recherche (ANR) under grants SemiTopo (ANR-12-BS04-0007), IsoTop (ANR-10-BLAN-0419) and TopoDyn (ANR-14-ACHN-0031) and by the European Research Council grant MesoQMC (ERC-2010-StG\_20091028 257241).

\bibliographystyle{elsarticle-num-names}
\bibliography{biblioFloquet}

\begin{thebibliography}{40}
\expandafter\ifx\csname natexlab\endcsname\relax\def\natexlab#1{#1}\fi
\providecommand{\url}[1]{\texttt{#1}}
\providecommand{\href}[2]{#2}
\providecommand{\path}[1]{#1}
\providecommand{\DOIprefix}{doi:}
\providecommand{\ArXivprefix}{arXiv:}
\providecommand{\URLprefix}{URL: }
\providecommand{\Pubmedprefix}{pmid:}
\providecommand{\doi}[1]{\href{http://dx.doi.org/#1}{\path{#1}}}
\providecommand{\Pubmed}[1]{\href{pmid:#1}{\path{#1}}}
\providecommand{\bibinfo}[2]{#2}
\ifx\xfnm\relax \def\xfnm[#1]{\unskip,\space#1}\fi
\bibitem[{Inoue and Tanaka(2010)}]{Inoue:2010}
\bibinfo{author}{J.-I. Inoue}, \bibinfo{author}{A.~Tanaka},
\newblock Photoinduced Transition between Conventional and Topological
  Insulators in Two-Dimensional Electronic Systems,
\newblock \bibinfo{journal}{Phys. Rev. Lett.} \bibinfo{volume}{105}
  (\bibinfo{year}{2010}) \bibinfo{pages}{017401}.
  \DOIprefix\doi{10.1103/PhysRevLett.105.017401}.
\bibitem[{{Lindner} et~al.(2011){Lindner}, {Refael}, and
  {Galitski}}]{Lindner:2011}
\bibinfo{author}{N.~H. {Lindner}}, \bibinfo{author}{G.~{Refael}},
  \bibinfo{author}{V.~{Galitski}},
\newblock Floquet topological insulator in semiconductor quantum wells,
\newblock \bibinfo{journal}{Nature Physics} \bibinfo{volume}{7}
  (\bibinfo{year}{2011}) \bibinfo{pages}{490--495}.
  \DOIprefix\doi{10.1038/nphys1926}. \href{http://arxiv.org/abs/1008.1792}{{\tt
  arXiv:1008.1792}}.
\bibitem[{Kitagawa et~al.(2011)Kitagawa, Oka, Brataas, Fu, and
  Demler}]{Kitagawa:2011}
\bibinfo{author}{T.~Kitagawa}, \bibinfo{author}{T.~Oka},
  \bibinfo{author}{A.~Brataas}, \bibinfo{author}{L.~Fu},
  \bibinfo{author}{E.~Demler},
\newblock Transport properties of nonequilibrium systems under the application
  of light: Photoinduced quantum Hall insulators without Landau levels,
\newblock \bibinfo{journal}{Phys. Rev. B} \bibinfo{volume}{84}
  (\bibinfo{year}{2011}) \bibinfo{pages}{235108}.
  \DOIprefix\doi{10.1103/PhysRevB.84.235108}.
\bibitem[{Kitagawa et~al.(2010)Kitagawa, Berg, Rudner, and
  Demler}]{Kitagawa:2010}
\bibinfo{author}{T.~Kitagawa}, \bibinfo{author}{E.~Berg},
  \bibinfo{author}{M.~Rudner}, \bibinfo{author}{E.~Demler},
\newblock Topological characterization of periodically driven quantum systems,
\newblock \bibinfo{journal}{Phys. Rev. B} \bibinfo{volume}{82}
  (\bibinfo{year}{2010}) \bibinfo{pages}{235114}.
  \DOIprefix\doi{10.1103/PhysRevB.82.235114}.
  \href{http://arxiv.org/abs/1010.6126}{{\tt arXiv:1010.6126}}.
\bibitem[{Rudner et~al.(2013)Rudner, Lindner, Berg, and Levin}]{Rudner:2013}
\bibinfo{author}{M.~S. Rudner}, \bibinfo{author}{N.~H. Lindner},
  \bibinfo{author}{E.~Berg}, \bibinfo{author}{M.~Levin},
\newblock Anomalous Edge States and the Bulk-Edge Correspondence for
  Periodically Driven Two-Dimensional Systems,
\newblock \bibinfo{journal}{Phys. Rev. X} \bibinfo{volume}{3}
  (\bibinfo{year}{2013}) \bibinfo{pages}{031005}.
  \DOIprefix\doi{10.1103/PhysRevX.3.031005}.
  \href{http://arxiv.org/abs/1212.3324}{{\tt arXiv:1212.3324}}.
\bibitem[{Carpentier et~al.(2015)Carpentier, Delplace, Fruchart, and
  Gawedzki}]{Carpentier:2015}
\bibinfo{author}{D.~Carpentier}, \bibinfo{author}{P.~Delplace},
  \bibinfo{author}{M.~Fruchart}, \bibinfo{author}{K.~Gawedzki},
\newblock Topological index for periodically driven time-reversal invariant 2D
  systems,
\newblock \bibinfo{journal}{Phys. Rev. Lett.} \bibinfo{volume}{114}
  (\bibinfo{year}{2015}) \bibinfo{pages}{106806}.
  \DOIprefix\doi{10.1103/PhysRevLett.114.106806}.
\bibitem[{Fang et~al.(2012)Fang, Yu, and Fan}]{Fang:2012}
\bibinfo{author}{K.~Fang}, \bibinfo{author}{Z.~Yu}, \bibinfo{author}{S.~Fan},
\newblock Realizing effective magnetic field for photons by controlling the
  phase of dynamic modulation,
\newblock \bibinfo{journal}{Nature Photonics} \bibinfo{volume}{6}
  (\bibinfo{year}{2012}) \bibinfo{pages}{782}.
  \DOIprefix\doi{10.1038/nphoton.2012.236}.
\bibitem[{{Kitagawa} et~al.(2012){Kitagawa}, {Broome}, {Fedrizzi}, {Rudner},
  {Berg}, {Kassal}, {Aspuru-Guzik}, {Demler}, and {White}}]{Kitagawa:2012}
\bibinfo{author}{T.~{Kitagawa}}, \bibinfo{author}{M.~A. {Broome}},
  \bibinfo{author}{A.~{Fedrizzi}}, \bibinfo{author}{M.~S. {Rudner}},
  \bibinfo{author}{E.~{Berg}}, \bibinfo{author}{I.~{Kassal}},
  \bibinfo{author}{A.~{Aspuru-Guzik}}, \bibinfo{author}{E.~{Demler}},
  \bibinfo{author}{A.~G. {White}},
\newblock Observation of topologically protected bound states in photonic
  quantum walks,
\newblock \bibinfo{journal}{Nature Communications} \bibinfo{volume}{3}
  (\bibinfo{year}{2012}). \DOIprefix\doi{10.1038/ncomms1872}.
  \href{http://arxiv.org/abs/1105.5334}{{\tt arXiv:1105.5334}}.
\bibitem[{Hauke et~al.(2012)Hauke, Tieleman, Celi, {\"O}lschl{\"a}ger, Simonet,
  Struck, Weinberg, Windpassinger, Sengstock, Lewenstein, and
  Eckardt}]{Hauke:2012}
\bibinfo{author}{P.~Hauke}, \bibinfo{author}{O.~Tieleman},
  \bibinfo{author}{A.~Celi}, \bibinfo{author}{C.~{\"O}lschl{\"a}ger},
  \bibinfo{author}{J.~Simonet}, \bibinfo{author}{J.~Struck},
  \bibinfo{author}{M.~Weinberg}, \bibinfo{author}{P.~Windpassinger},
  \bibinfo{author}{K.~Sengstock}, \bibinfo{author}{M.~Lewenstein},
  \bibinfo{author}{A.~Eckardt},
\newblock Non-Abelian Gauge Fields and Topological Insulators in Shaken Optical
  Lattices,
\newblock \bibinfo{journal}{Phys. Rev. Lett.} \bibinfo{volume}{109}
  (\bibinfo{year}{2012}) \bibinfo{pages}{145301}.
  \DOIprefix\doi{10.1103/PhysRevLett.109.145301}.
\bibitem[{Rechtsman et~al.(2013)Rechtsman, Zeuner, Plotnik, Lumer, Podolsky,
  Dreisow, Nolte, Segev, and Szameit}]{Rechtsman:2013}
\bibinfo{author}{M.~C. Rechtsman}, \bibinfo{author}{J.~M. Zeuner},
  \bibinfo{author}{Y.~Plotnik}, \bibinfo{author}{Y.~Lumer},
  \bibinfo{author}{D.~Podolsky}, \bibinfo{author}{F.~Dreisow},
  \bibinfo{author}{S.~Nolte}, \bibinfo{author}{M.~Segev},
  \bibinfo{author}{A.~Szameit},
\newblock Photonic Floquet topological insulators,
\newblock \bibinfo{journal}{Nature} \bibinfo{volume}{496}
  (\bibinfo{year}{2013}) \bibinfo{pages}{196--200}.
  \DOIprefix\doi{10.1038/nature12066}.
  \href{http://arxiv.org/abs/1212.3146}{{\tt arXiv:1212.3146}}.
\bibitem[{Hu et~al.(2015)Hu, Pillay, Wu, Pasek, Shum, and Chong}]{Hu:2015}
\bibinfo{author}{W.~Hu}, \bibinfo{author}{J.~C. Pillay},
  \bibinfo{author}{K.~Wu}, \bibinfo{author}{M.~Pasek}, \bibinfo{author}{P.~P.
  Shum}, \bibinfo{author}{Y.~D. Chong},
\newblock Measurement of a Topological Edge Invariant in a Microwave Network,
\newblock \bibinfo{journal}{Phys. Rev. X} \bibinfo{volume}{5}
  (\bibinfo{year}{2015}) \bibinfo{pages}{011012}.
  \DOIprefix\doi{10.1103/PhysRevX.5.011012}.
\bibitem[{Karzig et~al.(2014)Karzig, Bardyn, Lindner, and Refael}]{Karzig:2014}
\bibinfo{author}{T.~Karzig}, \bibinfo{author}{C.-E. Bardyn},
  \bibinfo{author}{N.~Lindner}, \bibinfo{author}{G.~Refael}, Topological
  polaritons from quantum wells in photonic waveguides or microcavities,
  \bibinfo{year}{2014}. \href{http://arxiv.org/abs/1406.4156}{{\tt
  arXiv:1406.4156}}.
\bibitem[{Reichl and Mueller(2014)}]{Reichl:2014}
\bibinfo{author}{M.~D. Reichl}, \bibinfo{author}{E.~J. Mueller},
\newblock Floquet edge states with ultracold atoms,
\newblock \bibinfo{journal}{Phys. Rev. A} \bibinfo{volume}{89}
  (\bibinfo{year}{2014}) \bibinfo{pages}{063628}.
  \DOIprefix\doi{10.1103/PhysRevA.89.063628}.
\bibitem[{{Jotzu} et~al.(2014){Jotzu}, {Messer}, {Desbuquois}, {Lebrat},
  {Uehlinger}, {Greif}, and {Esslinger}}]{Jotzu:2014}
\bibinfo{author}{G.~{Jotzu}}, \bibinfo{author}{M.~{Messer}},
  \bibinfo{author}{R.~{Desbuquois}}, \bibinfo{author}{M.~{Lebrat}},
  \bibinfo{author}{T.~{Uehlinger}}, \bibinfo{author}{D.~{Greif}},
  \bibinfo{author}{T.~{Esslinger}}, Experimental realisation of the topological
  Haldane model, \bibinfo{year}{2014}.
  \href{http://arxiv.org/abs/1406.7874}{{\tt arXiv:1406.7874}}.
\bibitem[{Wang et~al.(2013)Wang, Steinberg, Jarillo-Herrero, and
  Gedik}]{Wang:2013}
\bibinfo{author}{K.~H. Wang}, \bibinfo{author}{H.~Steinberg},
  \bibinfo{author}{P.~Jarillo-Herrero}, \bibinfo{author}{N.~Gedik},
\newblock Observation of Floquet-Bloch States on the Surface of a Topological
  Insulator,
\newblock \bibinfo{journal}{Science} \bibinfo{volume}{342}
  (\bibinfo{year}{2013}) \bibinfo{pages}{453}.
\bibitem[{Onishi et~al.(2014)Onishi, Ren, Novak, Segawa, Ando, and
  Tanaka}]{Onishi:2014}
\bibinfo{author}{Y.~Onishi}, \bibinfo{author}{Z.~Ren},
  \bibinfo{author}{M.~Novak}, \bibinfo{author}{K.~Segawa},
  \bibinfo{author}{Y.~Ando}, \bibinfo{author}{K.~Tanaka}, Instantaneous Photon
  Drag Currents in Topological Insulators, \bibinfo{year}{2014}.
  \href{http://arxiv.org/abs/1403.2492}{{\tt arXiv:1403.2492}}.
\bibitem[{Lazarides et~al.(2014)Lazarides, Das, and Moessner}]{Lazarides:2014}
\bibinfo{author}{A.~Lazarides}, \bibinfo{author}{A.~Das},
  \bibinfo{author}{R.~Moessner},
\newblock Equilibrium states of generic quantum systems subject to periodic
  driving,
\newblock \bibinfo{journal}{Phys. Rev. E} \bibinfo{volume}{90}
  (\bibinfo{year}{2014}) \bibinfo{pages}{012110}.
  \DOIprefix\doi{10.1103/PhysRevE.90.012110}.
\bibitem[{Dehghani et~al.(2014)Dehghani, Oka, and Mitra}]{Dehghani:2014}
\bibinfo{author}{H.~Dehghani}, \bibinfo{author}{T.~Oka},
  \bibinfo{author}{A.~Mitra},
\newblock Dissipative Floquet Topological Systems,
\newblock \bibinfo{journal}{Phys. Rev. B} \bibinfo{volume}{90}
  (\bibinfo{year}{2014}) \bibinfo{pages}{195429}.
  \DOIprefix\doi{10.1103/PhysRevB.90.195429}.
\bibitem[{Iadecola et~al.(2015)Iadecola, Neupert, and Chamon}]{Lacedola:2015}
\bibinfo{author}{T.~Iadecola}, \bibinfo{author}{T.~Neupert},
  \bibinfo{author}{C.~Chamon},
\newblock Occupation of topological Floquet bands in open systems,
\newblock \bibinfo{journal}{Phys. Rev. B} \bibinfo{volume}{91}
  (\bibinfo{year}{2015}) \bibinfo{pages}{235133}.
  \DOIprefix\doi{10.1103/PhysRevB.91.235133}.
\bibitem[{Seetharam et~al.(2015)Seetharam, Bardyn, Lindner, Rudner, and
  Refael}]{Seetharam:2015}
\bibinfo{author}{K.~I. Seetharam}, \bibinfo{author}{C.-E. Bardyn},
  \bibinfo{author}{N.~H. Lindner}, \bibinfo{author}{M.~S. Rudner},
  \bibinfo{author}{G.~Refael}, Controlled Population of Floquet-Bloch States
  via Coupling to Bose and Fermi Baths, \bibinfo{year}{2015}.
  \href{http://arxiv.org/abs/1502.02664}{{\tt arXiv:1502.02664}}.
\bibitem[{Dehghani et~al.(2015)Dehghani, Oka, and Mitra}]{Dehghani:2015}
\bibinfo{author}{H.~Dehghani}, \bibinfo{author}{T.~Oka},
  \bibinfo{author}{A.~Mitra},
\newblock Out-of-equilibrium electrons and the Hall conductance of a Floquet
  topological insulator,
\newblock \bibinfo{journal}{Phys. Rev. B} \bibinfo{volume}{91}
  (\bibinfo{year}{2015}) \bibinfo{pages}{155422}.
  \DOIprefix\doi{10.1103/PhysRevB.91.155422}.
\bibitem[{{B\"uttiker}(1988)}]{Buttiker:1988}
\bibinfo{author}{M.~{B\"uttiker}},
\newblock Absence of backscattering in the quantum Hall effect in multiprobe
  conductors,
\newblock \bibinfo{journal}{Phys. Rev. B} \bibinfo{volume}{38}
  (\bibinfo{year}{1988}) \bibinfo{pages}{9375}.
  \DOIprefix\doi{10.1103/PhysRevB.38.9375}.
\bibitem[{Roth et~al.(2009)Roth, Br{\"u}ne, Buhmann, Molenkamp, Maciejko, Qi,
  and Zhang}]{Roth:2009}
\bibinfo{author}{A.~Roth}, \bibinfo{author}{C.~Br{\"u}ne},
  \bibinfo{author}{H.~Buhmann}, \bibinfo{author}{L.~W. Molenkamp},
  \bibinfo{author}{J.~Maciejko}, \bibinfo{author}{X.-L. Qi},
  \bibinfo{author}{S.-C. Zhang},
\newblock Nonlocal Transport in the Quantum Spin {H}all State,
\newblock \bibinfo{journal}{Science} \bibinfo{volume}{325}
  (\bibinfo{year}{2009}) \bibinfo{pages}{294}.
  \DOIprefix\doi{10.1126/science.1174736}.
\bibitem[{Büttiker(2009)}]{buttiker2009}
\bibinfo{author}{M.~Büttiker},
\newblock Edge-State Physics Without Magnetic Fields,
\newblock \bibinfo{journal}{Science} \bibinfo{volume}{325}
  (\bibinfo{year}{2009}) \bibinfo{pages}{278--279}.
  \DOIprefix\doi{10.1126/science.1177157}.
\bibitem[{Gu et~al.(2011)Gu, Fertig, Arovas, and Auerbach}]{Gu:2011}
\bibinfo{author}{Z.~Gu}, \bibinfo{author}{H.~A. Fertig}, \bibinfo{author}{D.~P.
  Arovas}, \bibinfo{author}{A.~Auerbach},
\newblock Floquet Spectrum and Transport through an Irradiated Graphene Ribbon,
\newblock \bibinfo{journal}{Phys. Rev. Lett.} \bibinfo{volume}{107}
  (\bibinfo{year}{2011}) \bibinfo{pages}{216601}.
  \DOIprefix\doi{10.1103/PhysRevLett.107.216601}.
\bibitem[{Kundu and Seradjeh(2013)}]{Kundu:2013}
\bibinfo{author}{A.~Kundu}, \bibinfo{author}{B.~Seradjeh},
\newblock Transport Signatures of Floquet Majorana Fermions in Driven
  Topological Superconductors,
\newblock \bibinfo{journal}{Phys. Rev. Lett.} \bibinfo{volume}{111}
  (\bibinfo{year}{2013}) \bibinfo{pages}{136402}.
  \DOIprefix\doi{10.1103/PhysRevLett.111.136402}.
\bibitem[{Torres et~al.(2014)Torres, Perez-Piskunow, Balseiro, and
  Usaj}]{FoaTorres:2014}
\bibinfo{author}{L.~E. F.~F. Torres}, \bibinfo{author}{P.~M. Perez-Piskunow},
  \bibinfo{author}{C.~A. Balseiro}, \bibinfo{author}{G.~Usaj},
\newblock Multiterminal Conductance of a Floquet Topological Insulator,
\newblock \bibinfo{journal}{Phys. Rev. Lett.} \bibinfo{volume}{113}
  (\bibinfo{year}{2014}) \bibinfo{pages}{266801}.
  \DOIprefix\doi{0.1103/PhysRevLett.113.266801}.
\bibitem[{Dahlhaus et~al.(2015)Dahlhaus, Fregoso, and Moore}]{dahlhaus2015}
\bibinfo{author}{J.~P. Dahlhaus}, \bibinfo{author}{B.~M. Fregoso},
  \bibinfo{author}{J.~E. Moore},
\newblock Magnetization Signatures of Light-Induced Quantum Hall Edge States,
\newblock \bibinfo{journal}{Phys. Rev. Lett.} \bibinfo{volume}{114}
  (\bibinfo{year}{2015}) \bibinfo{pages}{246802}. \URLprefix
  \url{http://link.aps.org/doi/10.1103/PhysRevLett.114.246802}.
  \DOIprefix\doi{10.1103/PhysRevLett.114.246802}.
\bibitem[{Fregoso et~al.(2014)Fregoso, Dahlhaus, and Moore}]{fregoso2014}
\bibinfo{author}{B.~M. Fregoso}, \bibinfo{author}{J.~P. Dahlhaus},
  \bibinfo{author}{J.~E. Moore},
\newblock Dynamics of tunneling into nonequilibrium edge states,
\newblock \bibinfo{journal}{Phys. Rev. B} \bibinfo{volume}{90}
  (\bibinfo{year}{2014}) \bibinfo{pages}{155127}. \URLprefix
  \url{http://link.aps.org/doi/10.1103/PhysRevB.90.155127}.
  \DOIprefix\doi{10.1103/PhysRevB.90.155127}.
\bibitem[{Farrell and Pereg-Barnea(2015)}]{Farrell:2015}
\bibinfo{author}{A.~Farrell}, \bibinfo{author}{T.~Pereg-Barnea}, Edge State
  Transport in Floquet Topological Insulators, \bibinfo{year}{2015}.
  \href{http://arxiv.org/abs/1505.05584}{{\tt arXiv:1505.05584}}.
\bibitem[{{Kohler} et~al.(2005){Kohler}, {Lehmann}, and {Hanggi}}]{Kohler:2005}
\bibinfo{author}{S.~{Kohler}}, \bibinfo{author}{J.~{Lehmann}},
  \bibinfo{author}{P.~{Hanggi}},
\newblock {Driven quantum transport on the nanoscale},
\newblock \bibinfo{journal}{Physics Reports} \bibinfo{volume}{406}
  (\bibinfo{year}{2005}) \bibinfo{pages}{379--443}.
  \DOIprefix\doi{10.1016/j.physrep.2004.11.002}.
\bibitem[{Sambe(1973)}]{sambe73}
\bibinfo{author}{H.~Sambe},
\newblock Steady States and Quasienergies of a Quantum-Mechanical System in an
  Oscillating Field,
\newblock \bibinfo{journal}{Phys. Rev. A} \bibinfo{volume}{7}
  (\bibinfo{year}{1973}) \bibinfo{pages}{2203--2213}.
  \DOIprefix\doi{10.1103/PhysRevA.7.2203}.
\bibitem[{Moskalets and {B\"uttiker}(2002)}]{Moskalets:2002}
\bibinfo{author}{M.~Moskalets}, \bibinfo{author}{M.~{B\"uttiker}},
\newblock Floquet scattering theory of quantum pumps,
\newblock \bibinfo{journal}{Phys. Rev. B} \bibinfo{volume}{66}
  (\bibinfo{year}{2002}) \bibinfo{pages}{205320}.
  \DOIprefix\doi{10.1103/PhysRevB.66.205320}.
\bibitem[{Stefanucci et~al.(2008)Stefanucci, Kurth, Rubio, and
  Gross}]{Stefanucci:2008}
\bibinfo{author}{G.~Stefanucci}, \bibinfo{author}{S.~Kurth},
  \bibinfo{author}{A.~Rubio}, \bibinfo{author}{E.~K.~U. Gross},
\newblock Time-dependent approach to electron pumping in open quantum systems,
\newblock \bibinfo{journal}{Phys. Rev. B} \bibinfo{volume}{77}
  (\bibinfo{year}{2008}) \bibinfo{pages}{075339}.
  \DOIprefix\doi{10.1103/PhysRevB.77.075339}.
\bibitem[{Gaury et~al.(2013)Gaury, Weston, Santin, Houzet, Groth, and
  Waintal}]{Gaury:2013}
\bibinfo{author}{B.~Gaury}, \bibinfo{author}{J.~Weston},
  \bibinfo{author}{M.~Santin}, \bibinfo{author}{M.~Houzet},
  \bibinfo{author}{C.~Groth}, \bibinfo{author}{X.~Waintal},
\newblock Numerical simulations of time-resolved quantum electronics,
\newblock \bibinfo{journal}{Physics Reports} \bibinfo{volume}{534}
  (\bibinfo{year}{2013}) \bibinfo{pages}{1}.
  \DOIprefix\doi{10.1016/j.physrep.2013.09.001}.
\bibitem[{Fisher and Lee(1981)}]{Fisher:1981}
\bibinfo{author}{D.~S. Fisher}, \bibinfo{author}{P.~A. Lee},
\newblock Relation between conductivity and transmission matrix,
\newblock \bibinfo{journal}{Phys. Rev. B} \bibinfo{volume}{23}
  (\bibinfo{year}{1981}) \bibinfo{pages}{6851(R)}.
  \DOIprefix\doi{10.1103/PhysRevB.23.6851}.
\bibitem[{Arrachea and Moskalets(2006)}]{arrachea2006}
\bibinfo{author}{L.~Arrachea}, \bibinfo{author}{M.~Moskalets},
\newblock Relation between scattering-matrix and Keldysh formalisms for quantum
  transport driven by time-periodic fields,
\newblock \bibinfo{journal}{Phys. Rev. B} \bibinfo{volume}{74}
  (\bibinfo{year}{2006}) \bibinfo{pages}{245322}.
  \DOIprefix\doi{10.1103/PhysRevB.74.245322}.
\bibitem[{{Bernevig} et~al.(2006){Bernevig}, {Hughes}, and
  {Zhang}}]{Bernevig:2006}
\bibinfo{author}{B.~A. {Bernevig}}, \bibinfo{author}{T.~L. {Hughes}},
  \bibinfo{author}{S.-C. {Zhang}},
\newblock Quantum Spin Hall Effect and Topological Phase Transition in HgTe
  Quantum Wells,
\newblock \bibinfo{journal}{Science} \bibinfo{volume}{314}
  (\bibinfo{year}{2006}) \bibinfo{pages}{1757--1761}.
  \DOIprefix\doi{10.1126/science.1133734}.
  \href{http://arxiv.org/abs/cond-mat/0611399}{{\tt arXiv:cond-mat/0611399}}.
\bibitem[{Haldane(1988)}]{Haldane:1988}
\bibinfo{author}{F.~D.~M. Haldane},
\newblock Model for a Quantum {Hall} Effect without Landau Levels:
  Condensed-Matter Realization of the "Parity Anomaly",
\newblock \bibinfo{journal}{Phys. Rev. Lett.} \bibinfo{volume}{61}
  (\bibinfo{year}{1988}) \bibinfo{pages}{2015--2018}.
  \DOIprefix\doi{10.1103/PhysRevLett.61.2015}.
\bibitem[{Groth et~al.(2014)Groth, Wimmer, Akhmerov, and Waintal}]{Groth:2014}
\bibinfo{author}{C.~W. Groth}, \bibinfo{author}{M.~Wimmer},
  \bibinfo{author}{A.~R. Akhmerov}, \bibinfo{author}{X.~Waintal},
\newblock Kwant: a software package for quantum transport,
\newblock \bibinfo{journal}{New J. Phys} \bibinfo{volume}{16}
  (\bibinfo{year}{2014}) \bibinfo{pages}{063065}.
  \DOIprefix\doi{10.1088/1367-2630/16/6/063065}.

\end{thebibliography}

\appendix 
\section{Transmission coefficients}
\label{sec:appendix}

In the following, we extend the approach of \cite{Kohler:2005} to a quasi-unidimensional system in the geometry of Fig.~\ref{fig:figure_geometry}. The purpose is 
to describe the scattering through a two dimensional driven system,  viewed as a Chern Floquet filter.  We follow closely the derivation of \cite{Kohler:2005} and 
only highlight the differences due to the geometry considered. 

We consider a system composed of a central region $\mathcal{R}$ (the scattering region), which is submitted to a periodic excitation. This region is described by a time-periodic (with period $T=2\pi/\omega$) tight-binding Hamiltonian
\begin{equation}
	H_{\text{sys}} = \sum_{n n' \in \mathcal{R}} H_{n n'}(t) c_{n}^{\dagger} c_{n'}^{}
\end{equation}
with $H_{n n'}(t+T) = H_{n n'}(t)$, where $c_{n}$ is the annihilation operator of an electron in an localized state $n$ of the tight-binding model, the index $n$ representing the position on the Bravais lattice as well as internal degrees of freedom (sublattice, orbital, spin, etc.). 
This central region is connected to $N_{\text{leads}}$ leads described by the Hamiltonian
\begin{equation}
	H_{\text{leads}} = \sum_{\ell = 1}^{N_{\text{leads}}} \sum_{q \alpha} \energy_{\ell q \alpha} c_{\ell q \alpha}^{\dagger} c_{\ell q \alpha}, 
\end{equation}
where $\alpha$ labels transverse modes of the semi-infinite lead and $q$ is the longitudinal momentum in this lead. 
We assume that the wavefunctions in the leads read 
\begin{equation}
  \psi_{\alpha q}(x,y) = \frac{1}{\sqrt{L}} \ee^{- \ii q x} \chi_{\alpha}(y), 
\end{equation}
and in particular that transverse modes $\ket{\chi}$ do not depend on $q$ and constitute an orthonormal basis of the transverse Hilbert space,. 
(Notice that this is not always the case, especially when a magnetic field is present \cite{Groth:2014}). 
The annihilation operator of a state localized at transverse position $y=y_{\ell}(n)$ in the interface with the lead ~$\ell$ is written as 
\begin{equation}
  c_{\ell, y=y_{\ell}(n)} = \sum_{q \alpha} c_{\ell q \alpha} \chi_{\ell \alpha}^*(n).
\end{equation}
The internal degrees of freedom in the leads can be taken into account, if needed, by considering more leads.

The contacts between the central region and the leads are described by the Hamiltonian
\begin{equation}
	H_{\text{contacts}} = \sum_{\ell = 1}^{N_{\text{leads}}} \sum_{n \in \mathcal{M}_{\ell}} \sum_{q} V_{\ell} c_{\ell, y=y_{\ell}(n)}^{\dagger} c_{n} + \text{h.c.}
\end{equation}
where~$\mathcal{M}_{\ell}$ describes the set of sites of the central region at the interface with lead~$\ell$. In terms of the transverse modes creation/annihilation operators, this Hamiltonian reads
\begin{equation}
	H_{\text{contacts}} = \sum_{\ell = 1}^{N_{\text{leads}}} \sum_{n \in \mathcal{M}_{\ell}} \sum_{q \alpha} V_{\ell} \chi_{\ell \alpha}^*(n) c_{\ell q \alpha}^{\dagger} c_{n} + \text{h.c.}.
\end{equation}

Using the approach of \cite{Kohler:2005} which amounts to describe the correlations of the driven system in terms of the equilibrium noise in the leads,  
 we obtain the equation \eqref{eq:Il_stat} of main text with transmission coefficients
\begin{equation*}
	T_{\ell \ell'}^{(p)}(\energy) = \!\!\!\!
	\sum_{\substack{ m, n \in \mathcal{I}_{\ell} \\ m',n' \in \mathcal{I}_{\ell'} }}
	(\gf_{m m'}^{(p)}(\energy))^* \Gamma^{\ell}_{m n}(\energy + p \hbar \omega) \gf_{n n'}^{(p)}(\energy) \Gamma^{\ell'}_{n' m'}(\energy), 
\end{equation*}
that can be written as a trace on the interfaces (Eq. \eqref{eq:coefficient_transmission_general} of main text),
\begin{equation*}
  T_{\ell \ell'}^{(p)}(\energy) = \text{Tr} \left[
  [\gf^{(p)}(\energy)]^\dagger \Gamma^{\ell}(\energy + p \hbar \omega) \gf^{(p)}(\energy) \Gamma^{\ell'}(\energy)
  \right]
\end{equation*}
where
\begin{equation}
	\Gamma_{\ell}(\energy) = 2 \pi \sum_{q \alpha}  \left|V_{\ell}\right|^2 \ket{\chi_{\alpha}}\!\bra{\chi_{\alpha}} \delta(\energy - \energy_{\ell q \alpha}).
\end{equation} 
This expression is in agreement with previous approaches based on different formalisms \cite{Stefanucci:2008,Gaury:2013,arrachea2006}.

\end{document}